\documentstyle[12pt,epsf]{article}
\newcommand{\beq}{\begin{equation}}
\newcommand{\eeq}{\end{equation}}
\begin{document}
\begin{center}
{\bf SCALING FOR THE COALESCENCE OF MICROFRACTURES BEFORE BREAKDOWN}
\end{center}
\noindent
S. ZAPPERI$^{1}$, P. RAY$^2$, 
H.E. STANLEY$^{1}$ AND A. VESPIGNANI$^3$

\noindent
$^1$ Center for Polymer Studies and Department of Physics,
 Boston University, Boston, MA 02215 ,USA
\noindent
$^2$The Institute of Mathematical Sciences,
        CIT Campus, Madras - 600 113, India
\noindent
$^3$ Instituut-Lorentz, University of Leiden, P.O. Box 9506
The Netherlands.

\subsection*{\small ABSTRACT}
We study the behavior of fracture in disordered
systems close to the breakdown point. We simulate numerically 
both scalar (resistor network) and vectorial (spring network)
models with threshold disorder, driven at constant 
current and stress rate respectively.
We analyze the scaling of the susceptibility and the cluster
size close to the breakdown. We observe avalanche
behavior and clustering of the cracks. We find 
that the scaling exponents are consistent with those found
close to a mean-field spinodal and present
analogies between the coalescence of microfractures 
and the coalescence of droplets
in a metastable magnetic system.
Finally, we discuss different experimental conditions and
some possible theoretical interpretations of the results.
 
\subsection*{\small INTRODUCTION}

The breakdown of solids under external forces 
is a longstanding problem, that has practical and theoretical
relevance. The way a material breaks, under the effect 
of an external electric field or under mechanical stress
are closely related problems, due to the formal similarities
in the underlying laws governing those phenomena.
The first theoretical approach to fracture mechanics dates
back to the twenties with the work of Griffith \cite{griff}, who 
formulated a theory of crack formation, which is similar to 
the classical theory of nucleation in first-order phase
transitions. Cracks grow or heal, depending on whether
the external stress prevail over the resistance at surface
of the crack. Similarly in bubble nucleation \cite{nucl}, a critical
droplet will form when the change in free energy due 
to the bulk exceeds that of the surface.
Griffith theory assume the presence
a single microcrack of a particular shape surrounded
by an homogeneous medium, and therefore 
is not appropriate in disordered systems, where
cracks can start from different positions and coalescence
may take place.

Spinodal nucleation \cite{sn}, contrary 
to classical nucleation, is characterized by scaling
properties and fractal droplets. The spinodal point
in fact has some characteristics of a critical point in second
order phase transitions.
The similarity between a solid driven to the threshold
of mechanical instability and spinodal nucleation
has been discussed in the past. Rundle and Klein 
\cite{rk}, using a Landau-Ginzburg
analysis of a single crack, showed that the crack growth
obeyed scaling laws expected for spinodal nucleation.
Selinger et al. \cite{sel1} have 
shown by numerical simulations and mean-field theory
of thermally activated fracture that the 
breakdown has the characteristics 
of a spinodal point. 

In this paper we concentrate on
disordered media and we disregard the effect thermal
fluctuations. The system is driven by an increasing
external load to the point of global failure. It has been 
experimentally observed that the response, 
detected by acoustic emission (AE) measurements, 
to an increasing external stress takes place in bursts or avalanches
distributed over a wide range of scale. Examples of this
are found in foam glasses \cite{pith},
fiber matrix composites \cite{sorn3}, concretes \cite{ae}, 
hydrogen precipitation \cite{ccc} and volcanic rocks \cite{dmp}.
We observe a similar behavior for two dimensional discrete
models. We show that the scaling behavior close to the 
breakdown is in quantitative agreement with the mean-field theory
and it is suggestive of a first-order transition. 
The values of the scaling exponents are 
consistent with those found close
to a spinodal point in {\em thermally} driven homogeneous
systems. 

\subsection*{\small THE MODELS}

We study here two models, the random fuse model \cite{fuse}
for electric breakdown and a spring network model \cite{spring}
for fracture.
In the random fuse model \cite{fuse} each bond of a two dimensional
lattice is occupied by a fuse of conductivity $\sigma=1$,
which burns when the current flowing in it exceeds a 
quenched random threshold. We consider a rotated square
lattice with periodic boundary conditions in one direction.
We impose a constant external current on the two
other edges of the lattice. The currents
in each bond are computed solving the kirchhoff equations.
This step corresponds to the minimization of the total energy
dissipated in the lattice
\beq
E(\{\sigma\})\equiv\frac{1}{2}\sum_i \sigma_i (\Delta V)_i^2,
\label{en}
\eeq
where $(\Delta V)_i$ is the voltage drop in the bond $i$.
We employ a multigrid relaxation algorithm 
with precision $\epsilon = 10^ {-10}$. When all the currents are
below the threshold we increase the current until 
the next bond reaches the threshold. The process is
continued until a path of broken bonds spans the lattice
and no current flows anymore. We chose a uniform distribution
of thresholds, $D \in [0,2]$.

The second model is an elastic network \cite{spring} which has central and
rotationally invariant bond-bending forces. The potential energy is
\beq
E=\frac{a}{2}\sum_{i}(\delta r_{i})^2 \sigma_{i}+
\frac{b}{2}\sum_{<i,j>}(\delta\theta_{ij})^2 \sigma_{i}\sigma_{j}
\eeq
where $\delta r_{i}$ is the change in the length of the
bond $i$ and $\delta\theta_{ij}$ is the change in the 
angle between neighboring bonds $i$ and $j$. The constant
$\sigma_i$ is equal to one if the bond is present and it is zero
otherwise.
A slowly increasing external stress
is applied on all the edges and the lattice 
dynamics is obtained by numerically solving the equations of
motion for each spring. Bonds break when stretched beyond
a randomly chosen threshold.

\subsection*{\small SIMULATION RESULTS}

The response of the model to the increase of the external
force takes place in widely distributed avalanches.
The average size of the avalanches (i.e. the number of
broken bonds) increases when the global
failure is approached. We were able to show \cite{long} using mean-field
theory that the average avalanche size $\langle m\rangle$ diverges at
the breakdown as 
\beq
\langle m \rangle \sim (f_c-f)^{-\gamma}~~~~~~\gamma=1/2.
\label{susc}
\eeq
where $f$ is the stress or the current per unit length
imposed on the lattice. We note that the same scaling law
is expected close to a spinodal point, in the case of
thermally driven first-order transitions. The macroscopic
quantities of the system (i.e. elasticity) have
a finite jump at the breakdown, indicative of a first-order
transition.

\begin{figure}[htb]
\centerline{
        \epsfxsize=7.0cm
        \epsfbox{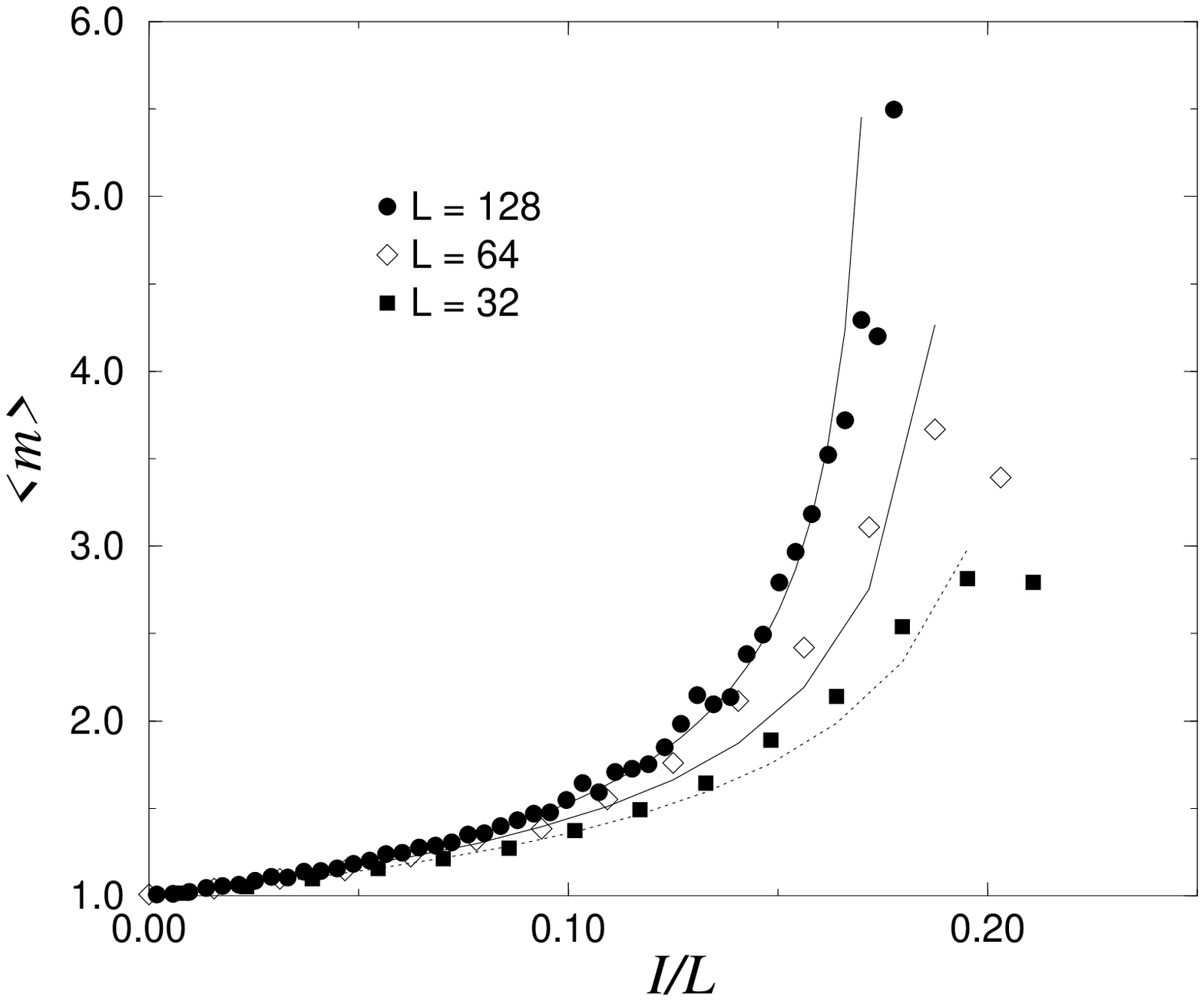}
        \vspace*{0.5cm}
        }
\centerline{
        \epsfxsize=7.0cm
        \epsfbox{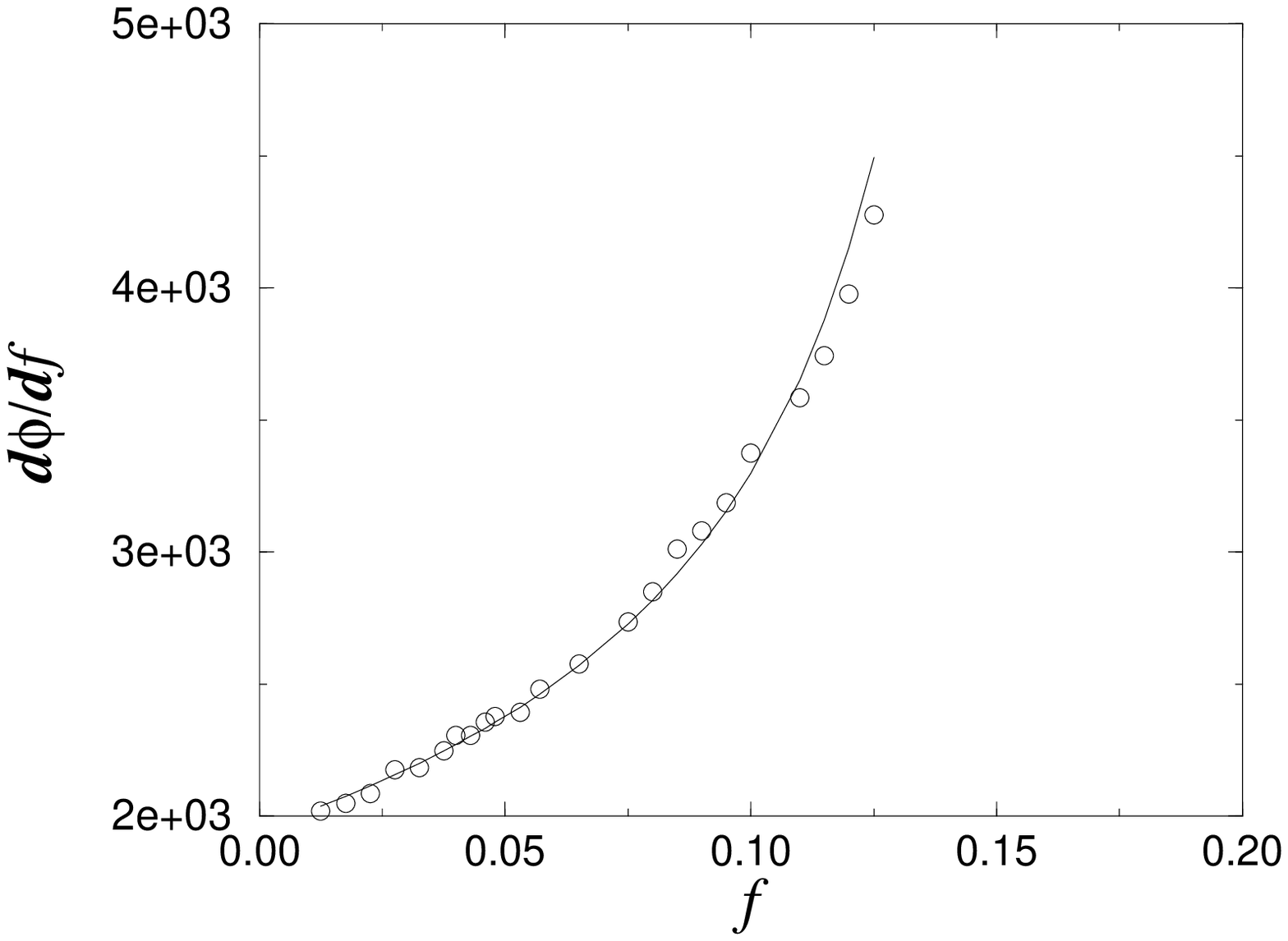}
        \vspace*{0.5cm}
        }
\caption{a) The average avalanche size $\langle m \rangle$
is plotted as a function of $f=I/L$, the fit is done 
using the mean-field value $\gamma=1/2$. b) The ``susceptibility''
of the spring network with the mean field fit ($\gamma=1/2$).
The parameter $\phi$ is the fraction of bond that are not
broken. The average avalanche size $\langle m \rangle$ is proportional
to $d\phi/df$.}
\label{fig1}
\end{figure}

We confirm the validity of mean-field 
scaling by computer simulations of two dimensional models. 
For both models mean-field theory
is obeyed remarkably well (see Fig.~\ref{fig1}a
and Fig.~\ref{fig1}b). 

\begin{figure}[htb]
\centerline{
        \epsfxsize=7.0cm
        \epsfbox{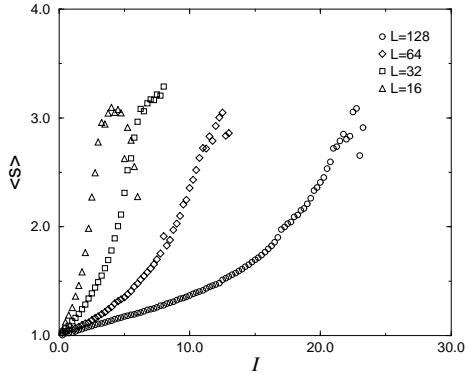}
        \vspace*{0.5cm}
        }
\caption{The average crack size for the fuse model as a function of
the current for different systems sizes. 
The crack size does not seem to diverge at the breakdown.}
\label{fig3}
\end{figure}

The reason for the observed mean-field behavior is probably
due to the long-range nature of elastic interactions.
The formation of cracks in those models takes place 
by the coalescence of several microcracks. This is 
confirmed by the behavior of the average crack size
which {\em does not diverge} at the breakdown (see Fig~\ref{fig3}).

\subsection*{\small DISCUSSION AND CONCLUSIONS}

The breakdown of driven disordered media is described
by scaling law which are reminiscent of those found close to
a spinodal point. It appears that the behavior
of a driven disordered system is similar to that of a thermally
driven homogeneous system. This analogy is not too strict since
the concept of metastability and spinodal are not well defined
in the first case.

Despite several experimental investigations of
avalanche dynamics in fractures \cite{pith,sorn3,ae,ccc,dmp}, 
there is not a clear theoretical interpretation of the results.
We believe that different experimental
conditions can all give rise to similar scaling behavior, 
but the underlying physical mechanisms could be quite different. 
We can distinguish the following experimental setups:
\begin{enumerate}
\item A solid driven by a constant stress rate
can be described in the framework discussed in this paper. 
The system responds to the increase of the external load by
AE bursts of increasing size \cite{sorn3}, diverging
at the point of global failure. It would be interesting
to check if the scaling exponents agree with the mean-field
theory.
\item A solid subject to a constant load breaks because
of thermal fluctuations. The AE is due to the formation
of ``droplets'' and should be power law distributed close to
the limit of stability (spinodal). Scaling exponents 
consistent with those of spinodal nucleation were
observed in a recent experiment on 
cellular glass \cite{pith}. To confirm this interpretation it would
be necessary to study the scaling for different values of
the applied load.
\item A solid in a perfectly plastic state 
could respond to the increase of the external strain
by a {\em stationary} AE signal. In this case one
can interpret the results as a manifestation of self-organized
criticality. Such a behavior was shown in numerical 
models \cite{zvs,gran}, but to our knowledge 
it has not yet been observed in experiments.
\end{enumerate}
We believe that  extensive and systematic experiments
along these lines can resolve these longstanding problems.

\subsection*{\small ACKNOWLEDGMENTS}
We wish to thank P. Cizeau, W. Klein, F. Sciortino and H. Strauven 
for useful discussions and remarks. The Center for Polymer studies is
supported by NSF.

\end{document}